\documentclass[10pt,conference]{IEEEtran}
\usepackage{cite,graphicx,psfrag,amsmath,amssymb,subfigure,url}
\newtheorem{theorem}{Theorem}

\def\CampCost{L}
\def\definedas{\triangleq}
\def\s{\mbox{'s}}
\def\boldp{p}
\def\bigp{P}
\def\boldw{w}
\def\bigw{W}
\def\kval{k}
\def\len{n}
\def\biglen{N}

\def\Rp{{{\mathbb R}_+}}

\def\X{{\mathcal X}}

\def\lg{{\log_2}}
\def\papertype{paper }

\newcommand{\defn}[0]{\it}

\begin{document}
\title{Infinite-Alphabet Prefix Codes Optimal for $\beta$-Exponential
Penalties}
\author{\authorblockN{Michael B. Baer}
\authorblockA{Electronics for Imaging\\
303 Velocity Way\\
Foster City, California  94404  USA\\
Email: Michael.Baer@efi.com}}

\maketitle

\begin{abstract}
Let $\bigp = \{\boldp(i)\}$ be a measure of strictly positive
probabilities on the set of nonnegative integers.  Although the
countable number of inputs prevents usage of the Huffman algorithm,
there are nontrivial $\bigp$ for which known methods find a source
code that is optimal in the sense of minimizing expected codeword
length.  For some applications, however, a source code should instead
minimize one of a family of nonlinear objective functions,
$\beta$-exponential means, those of the form $\log_a \sum_i \boldp(i)
a^{\len(i)}$, where $\len(i)$ is the length of the $i$th codeword and
$a$ is a positive constant.  Applications of such minimizations
include a problem of maximizing the chance of message receipt in
single-shot communications ($a<1$) and a problem of minimizing the
chance of buffer overflow in a queueing system ($a>1$).  This
\papertype introduces methods for finding codes optimal for such
exponential means.  One method applies to geometric distributions,
while another applies to distributions with lighter tails.  The latter
algorithm is applied to Poisson distributions.  Both are extended to
minimizing maximum pointwise redundancy.
\end{abstract}

\section{Introduction, Motivation, and Main Results} 
\label{intro} 

If probabilities are known, optimal lossless source coding of
individual symbols (and blocks of symbols) is usually done using David
Huffman's famous algorithm\cite{Huff}.  There are, however, cases that
this algorithm does not solve\cite{Abr01}.  Problems with an
infinite number of possible inputs --- e.g., geometrically-distributed
variables --- are not covered.  Also, in some instances, the
optimality criterion --- or {\defn penalty} --- is not the linear
penalty of expected length.  Both variants of the problem have been
considered in the literature, but not simultaneously.  This \papertype 
discusses cases which are both infinite and nonlinear.

An infinite-alphabet source emits symbols drawn from the alphabet
$\X_\infty = \{0, 1, 2, \ldots \}$.  (More generally, we use $\X$ to
denote an input alphabet whether infinite or finite.)  Let $\bigp =
\{\boldp(i)\}$ be the sequence of probabilities for each symbol, so
that the probability of symbol $i$ is $\boldp(i) > 0$.  The source
symbols are coded into binary codewords.  The codeword $c(i) \in
\{0,1\}^*$ in code $C$, corresponding to input symbol $i$, has length
$\len(i)$, thus defining length distribution $\biglen$.  

Perhaps the most well-known such codes are the optimal codes derived
by Golomb for geometric distributions\cite{Golo,GaVV}.  There are many
reasons for using infinite-alphabet codes rather than codes for finite
alphabets, such as Huffman codes.  The most obvious use is for cases
with no upper bound --- or at least no known upper bound --- on the
number of possible items.  In addition, for many cases it is far
easier to come up with a general code for integers rather than a
Huffman code for a large but finite number of inputs.  Similarly, it
is often faster to encode and decode such well-structured codes.  For
these reasons, infinite-alphabet codes and variants of them are widely
used in image and video compression standards\cite{WSBL, WSS}, as well
as for compressing text, audio, and numerical data.

To date, the literature on infinite-alphabet codes has considered only finding
efficient uniquely decipherable codes with respect to minimizing
expected codeword length $\sum_i \boldp(i) \len(i)$.  Other utility
functions, however, have been considered for finite-alphabet codes.
Campbell~\cite{Camp} introduced a problem in which the penalty to
minimize, given some continuous (strictly) monotonic increasing {\defn
cost function} $\varphi(x):\Rp \rightarrow \Rp$, is 
$$
\CampCost(\bigp,\biglen,\varphi) = \varphi^{-1}\left(\sum_i \boldp(i)
\varphi(\len(i))\right) 
$$ and specifically considered the exponential subcases with exponent $a>1$:
\begin{equation} 
\CampCost_a(\bigp,\biglen) \definedas \log_a \sum_i \boldp(i) a^{\len(i)} , 
\label{ExpCost} 
\end{equation} 
that is, $\varphi(x) = a^x$.  Note that minimizing penalty $\CampCost$
is also an interesting problem for $0<a<1$ and approaches the standard
penalty $\sum_i \boldp(i) \len(i)$ for $a \rightarrow 1$\cite{Camp}.
While $\varphi(x)$ decreases for $a<1$, one can map decreasing
$\varphi$ to a corresponding increasing function $\tilde{\varphi}(l) =
\varphi_{\max} - \varphi(l)$ (e.g., for $\varphi_{\max} = 1$) without
changing the penalty value.  Thus this problem, equivalent to
maximizing $\sum_i \boldp(i) a^{\len(i)}$, is a subset of those
considered by Campbell.  All penalties of the form (\ref{ExpCost}) are
called $\beta$-exponential means, where $\beta = \lg a$\cite[p.~158]{AcDa}.

Campbell noted certain properties for $\beta$-exponential means, but
did not consider applications for these means.  Applications were
later found for the problem with $a>1$\cite{Jeli,Humb2,BlMc}.  These
applications relate to a problem in which we wish to minimize the
probability of buffer overflow in communications; this is discussed in
the full version of this paper\cite{Baer07}.  Also discussed in the
full version is an application for $a<1$ introduced in \cite{BaerI06},
a problem of maximizing the chance of message receipt in single-shot
communications.

One can solve any instance of the exponential penalty with a finite
number of inputs using a linear-time algorithm found independently by
Hu {\it et al.}~\cite[p.~254]{HKT}, Parker \cite[p.~485]{Park0}, and
Humblet \cite[p.~25]{Humb0},\cite[p.~231]{Humb2}, although only the
last of these considered $a < 1$.  We present the exponential-penalty
algorithm here; even though it cannot be used for an infinite
alphabet, it can be used to derive and show the optimality of
infinite-alphabet codes:

{\bf Procedure for Exponential Huffman Coding} 

This procedure minimizes (\ref{ExpCost}) for any positive $a \neq 1$
and $|\X|<\infty$, even if the ``probabilities'' do not add to $1$.
We refer to such arbitrary positive inputs as {\defn weights}, denoted
by $\boldw(i)$ instead of~$\boldp(i)$:

\begin{enumerate}
\item Each item $i$ has weight $\boldw(i) \in \bigw_{\X}$, where $\X$
is the (finite) alphabet and $\bigw_{\X} = \{w(i)\}$ is the set of all
such weights.  Assume each item $i$ has codeword $c(i)$, to be
determined later.
\item Combine the items with the two smallest weights $\boldw(j)$ and
  $\boldw(k)$ into one item with the combined weight
  $\tilde{\boldw}(j) = a \cdot (\boldw(j) + \boldw(k))$.  This item
  has codeword $\tilde{c}(j)$, to be determined later, while item $j$ is
  assigned codeword $c(j) = \tilde{c}(j)0$ and $k$ codeword $c(k) =
  \tilde{c}(j)1$.  Since these have been assigned in terms of
  $\tilde{c}(j)$, replace $\boldw(j)$ and $\boldw(k)$ with
  $\tilde{\boldw}(j)$ in $\bigw_\X$ to form $\bigw_{\tilde{\X}}$.
\item Repeat procedure, now with the remaining codewords (reduced in
  number by $1$) and corresponding weights, until only one
  item is left.  The weight of this item is $\sum_i \boldw(i)
  a^{\len(i)}$.  All codewords are now defined by assigning the null
  string to this trivial item.
\end{enumerate}

Optimality of the algorithm is justified as in Huffman coding, in that
an exchange argument can be used to show that an optimal code exists
for which the least likely two codewords differ in only their final
bit, allowing a reduction to the equivalent smaller problem that linearly
combines their weights.  This
algorithm can be modified to run in linear time (to input size) given
sorted weights in the same manner as Huffman coding~\cite{Leeu}.  

Note that this algorithm assigns an explicit weight to each node of
the resulting code tree implied by having each item represented by a
node with its parent representing the combined items: If a node is a
leaf, its weight is given by the associated probability; otherwise its
weight is defined recursively as $a$ times the sum of its children.
This concept is useful in visualizing both the coding procedure
and its output.

It is also worthwhile to note that $a \leq 0.5$ is degenerate, always
resulting in the {\defn unary code} (for infinite inputs) or a
unary-like code (for finite inputs) being optimal for any probability
distribution.  The unary code has ones terminated by a zero, i.e.,
codewords of the form $\{1^i0 : i \geq 0\}$.  The unary-like code is a
truncated unary code, that is, a code with identical codewords to the
unary code except for the longest codeword, which is of the form
$1^{|\X|-1}$.  For the unary-like code, optimality for $a \leq 0.5$
can be shown using the coding procedure; the smallest two items, $j$
and $k$, are combined, and the resulting item has weight $a \cdot
(w(j)+w(k))$.  This is no larger than the larger of the constituent
weights, meaning that the resulting item will be combined with
third-smallest item, and so forth, resulting in a unary-like code.
Taking limits, informally speaking, results in a unary limit code;
formally, this is a straightforward corollary of Theorem~\ref{tailthm}
in Section~\ref{other}.

If $a>0.5$, a code with finite penalty exists if and only if R\'{e}nyi entropy 
of order $\alpha(a) = {(1+\lg a)}^{-1}$ is finite\cite{Baer06}.  It
was Campbell who first noted the connection between the optimal code's penalty,
$\CampCost_a(\bigp,\biglen^*)$, and R\'{e}nyi entropy
\begin{equation*}
\begin{array}{rcl}
H_{\alpha}(\bigp) &\definedas& \frac{1}{1-\alpha} \lg \sum_{i \in \X} 
\boldp(i)^\alpha \\
\Rightarrow H_{\alpha(a)}(\bigp) &=& \frac{1+\lg a}{\lg a} \lg \sum_{i \in \X} \boldp(i)^{(1+\lg a)^{-1}} .
\end{array}
\end{equation*}
This relationship is
$$H_{\alpha(a)}(\bigp) \leq \CampCost_a(\bigp,\biglen^*) <
H_{\alpha(a)}(\bigp) + 1$$ which should not be surprising given the
similar relationship between Huffman coding and Shannon entropy\cite{Shan},
which corresponds to $a \rightarrow 1$, $H_1(\bigp)$~\cite{Ren2}.

One must be careful regarding the meaning of an ``optimal code'' when
there are an infinite number of possible codes under consideration.
One might ask whether there must exist an optimal code or if there can
be an infinite sequence of codes of decreasing penalty without any
code achieving the limit penalty value.  Fortunately the answer is the
former, the proof being a special case of Theorem~2 in~\cite{Baer06}.
The question is then how to find one of these optimal source codes
given parameter $a$ and probability measure~$\bigp$.

As in the linear case, this is not known for general $\bigp$, but can
be found for certain common distributions.  In the next section, we
consider geometric distributions and find that Golomb codes are
optimal, although the optimal Golomb code for a given probability mass
function varies according to $a$.  The main result of
Section~\ref{geometric} is that, for $\boldp_\theta(i) =
(1-\theta)\theta^i$ and $a \in \Rp$, G$\kval$, the Golomb code
with parameter $k$, is optimal for $$\kval = \max\left(1, \left\lceil
-\log_\theta a -\log_\theta (1+\theta) \right\rceil\right).$$ In
Section~\ref{other}, we consider distributions that are relatively
light-tailed, that is, that decline faster than certain geometric
distributions.  If there is a nonnegative integer $r$ such that for
all $j>r$ and $i<j$,
$$\boldp(i) \geq \max\left(\boldp(j),\sum_{k=j+1}^\infty \boldp(k)
a^{k-j}\right)$$ then an optimal binary prefix code can be found which
is a generalization of the unary code.  A specific case of this is the
Poisson distribution, where an
aforementioned $r$ is given by $r = \max(\lceil 2 a \lambda \rceil -
2, \lceil e \lambda \rceil - 1)$ for $\boldp_\lambda(i)=\lambda^i
e^{-\lambda}/i!$.  Section~\ref{nonexp} discusses the maximum
pointwise redundancy penalty, which has a similar solution with
light-tailed distributions and for which the Golomb code G$\kval$ with
$\kval = \lceil -1/\lg \theta \rceil$ is optimal for with geometric
distributions.  Complete proofs and illustrations, as well as
additional results, are given in the full version\cite{Baer07}.

\section{Geometric Distribution with Exponential Penalty}
\label{geometric}

Consider the geometric distribution $$\boldp_\theta(i) =
(1-\theta)\theta^i$$ for parameter $\theta \in (0,1)$.  This
distribution arises in run-length coding as well as in other
circumstances\cite{Golo,GaVV}.

For the traditional linear penalty, a Golomb code with
parameter~$\kval$ --- or G$\kval$ --- is optimal for $\theta^\kval +
\theta^{\kval+1} \leq 1 < \theta^{\kval-1} + \theta^\kval$.  Such a
code consists of a unary prefix followed by a binary suffix, the
latter taking one of $\kval$ possible values.  If $\kval$ is a power
of two, all binary suffix possibilities have the same length;
otherwise, their lengths $\sigma(i)$ differ by at most $1$ and $\sum_i
2^{-\sigma(i)}=1$.  Such binary codes are called {\defn complete}
codes.  This defines the Golomb code; for example, the Golomb code for
$\kval = 3$ is:
\begin{center}
$$
\begin{array}{rll}
\hline
\hline
i&\boldp(i)&c(i) \\
\hline
0&1-\theta&0~0 \\
1&(1-\theta)\theta&0~10 \\
2&(1-\theta)\theta^2&0~11 \\
3&(1-\theta)\theta^3&10~0 \\
4&(1-\theta)\theta^4&10~10 \\
5&(1-\theta)\theta^5&10~11 \\
6&(1-\theta)\theta^6&110~0 \\
7&(1-\theta)\theta^7&110~10 \\
\vdots&\qquad \vdots&\qquad \vdots \\
\end{array}
$$
\end{center}
where the space in the code separates the unary prefix from the complete
suffix.  In general, codeword $j$ for G$\kval$ is of the form
$\{1^{\lfloor j/\kval \rfloor} 0 b(j \bmod \kval,\kval) : j \geq 0\}$,
where $b(j \bmod \kval, \kval)$ is a complete binary code for the $(j -
\kval \lfloor j/\kval \rfloor+1)$th of $\kval$ items.

It turns out that such codes are optimal for the exponential penalty:
\begin{theorem}
For $a \in \Rp$, if
\begin{equation}
\theta^\kval + \theta^{\kval+1} \leq \frac{1}{a} < \theta^{\kval-1} + \theta^\kval 
\label{ineq}
\end{equation}
for $\kval \geq 1$, then the Golomb code G$\kval$ is the optimal code
for $\bigp_\theta$.  If no such $\kval$ exists, the
unary code is optimal.  
\label{optgeo}
\end{theorem}

The proof of optimality (in full version \cite{Baer07}) uses the
procedure for exponential Huffman coding to find an optimal
exponential Huffman code for a sequence of similar finite weight
distributions.  Define an $m$-reduced geometric source $\bigw_m$ as:
$$
\boldw_m(i) \definedas \left\{
\begin{array}{ll}
(1-\theta)\theta^i,& 0 \leq i \leq m \\
\frac{(1-\theta)a\theta^i}{1-a\theta^\kval},& m < i \leq m + \kval \\
\end{array}
\right.
$$ for any $m \geq -1$. It can be shown that this distribution has an
optimal code with lengths $\len(0)$ through $\len(m)$ that are
identical to the Golomb code in question.  One can then show that the
optimal code for the geometric distribution must have a penalty
between that for the Golomb code for the geometric distribution and
the optimal code for $\bigw_m$ (for any $m$).  Since the latter two
penalties approach equality as $m \rightarrow \infty$, the Golomb code
must be optimal.

This rule for finding an optimal Golomb G$\kval$ code is equivalent to
$$\kval = \max\left(1, \left\lceil -\log_\theta a -\log_\theta
(1+\theta) \right\rceil\right).$$ This is a generalization of the
traditional linear result since this corresponds to $a \rightarrow 1$.
Cases in which the left inequality of (\ref{ineq}) is an equality have multiple
solutions, as with linear coding; see, e.g., \cite[p.~289]{Goli2}.

It is equivalent for the bits of the unary prefix to be reversed,
that is, to use $\{0^{\lfloor j/\kval \rfloor} 1 b(j \bmod
\kval,\kval) : j \geq 0\}$ (as in \cite{GaVV}) instead of
$\{1^{\lfloor j/\kval \rfloor} 0 b(j \bmod \kval,\kval) : j \geq 0\}$
(as in \cite{Golo}).  The latter has the advantage of being
alphabetic, that is, $i>j$ if and only if $c(i)$ is lexicographically
after $c(j)$.

A little algebra reveals that, for a distribution $\bigp_\theta$ and a Golomb
code with parameter $\kval$ (lengths $\biglen_\kval$), 
\begin{equation}
\begin{array}{rcl}
\CampCost_a(\bigp_\theta,\biglen_\kval) &=& \log_a \sum_{i=0}^\infty
(1-\theta)\theta^i a^{(\left\lceil\frac{i+1-z}{\kval} \right\rceil +
g)} \\ &=& g + {\log}_a
\left(1+\frac{(a-1)\theta^z}{1-a\theta^\kval}\right) 
\end{array}
\label{geosum}
\end{equation}
where
$g=\lfloor \log_2 \kval \rfloor + 1$ and $z = 2^g - \kval$.  
Therefore, Theorem~\ref{optgeo} provides the $\kval$ that minimizes
(\ref{geosum}).  If $a>0.5$, the corresponding R\'{e}nyi entropy is
\begin{equation}
H_{\alpha(a)}(\bigp_\theta) = \log_a
\frac{1-\theta}{(1-\theta^{\alpha(a)})^{1/\alpha(a)}}
\label{geoent}
\end{equation}
where we recall that $\alpha(a) = (1 +
\lg a)^{-1}$.  (Again, $a \leq 0.5$ is degenerate, an
optimal code being unary with no corresponding R\'{e}nyi entropy.)

\begin{figure*}[ht]
\psfrag{L-Ha}{$\bar{R}_a(\biglen_{\theta,a}^*,\bigp_\theta)$}
\psfrag{a}{$a$}
\psfrag{ag}{\mbox {\huge $a$}}
\psfrag{Theta}{$\theta$}
     \centering
     \subfigure[$a>1$]
	       { \label{apos} \includegraphics[width=.45\textwidth]{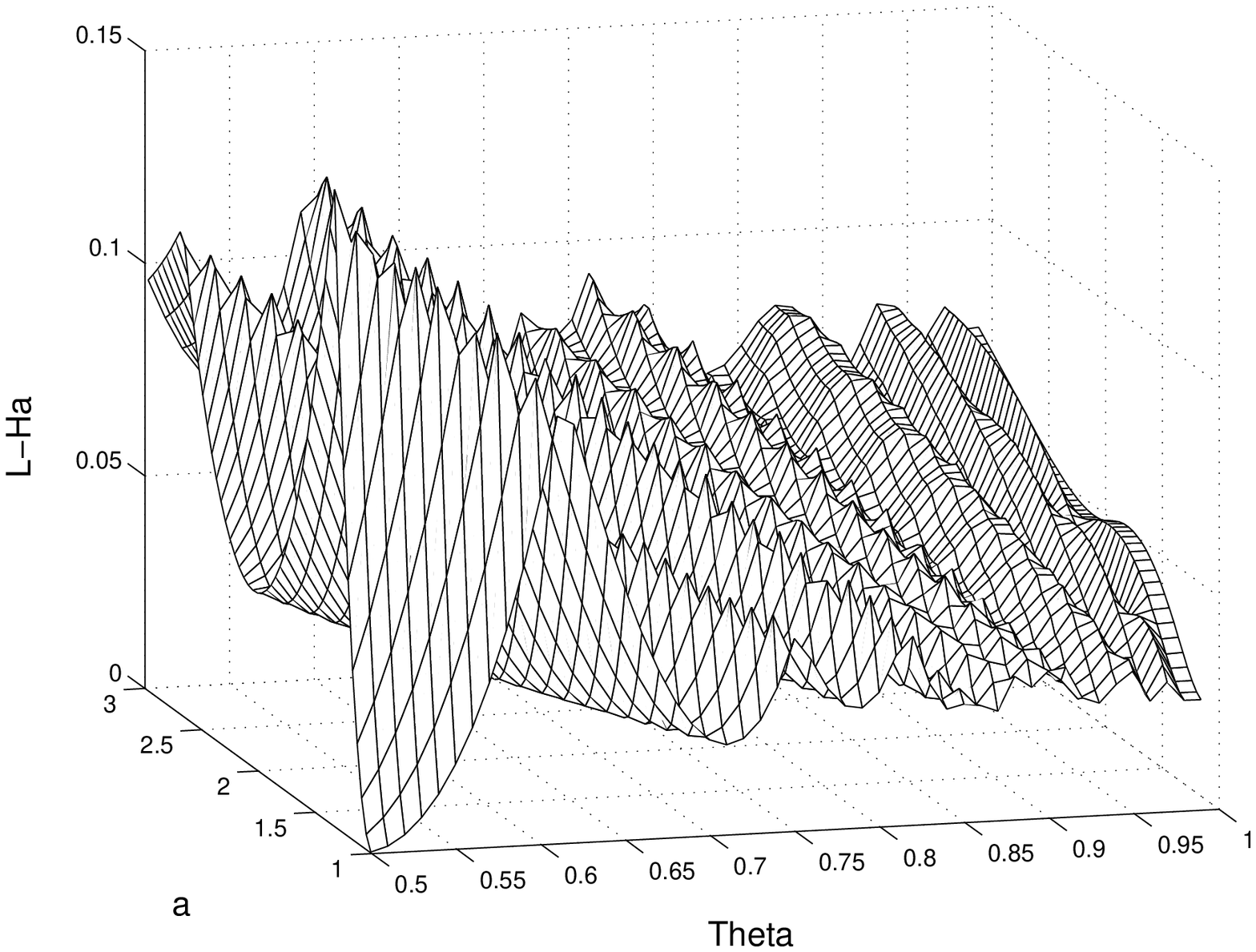} }
     \subfigure[$a<1$]
	       { \label{aneg} \includegraphics[width=.45\textwidth]{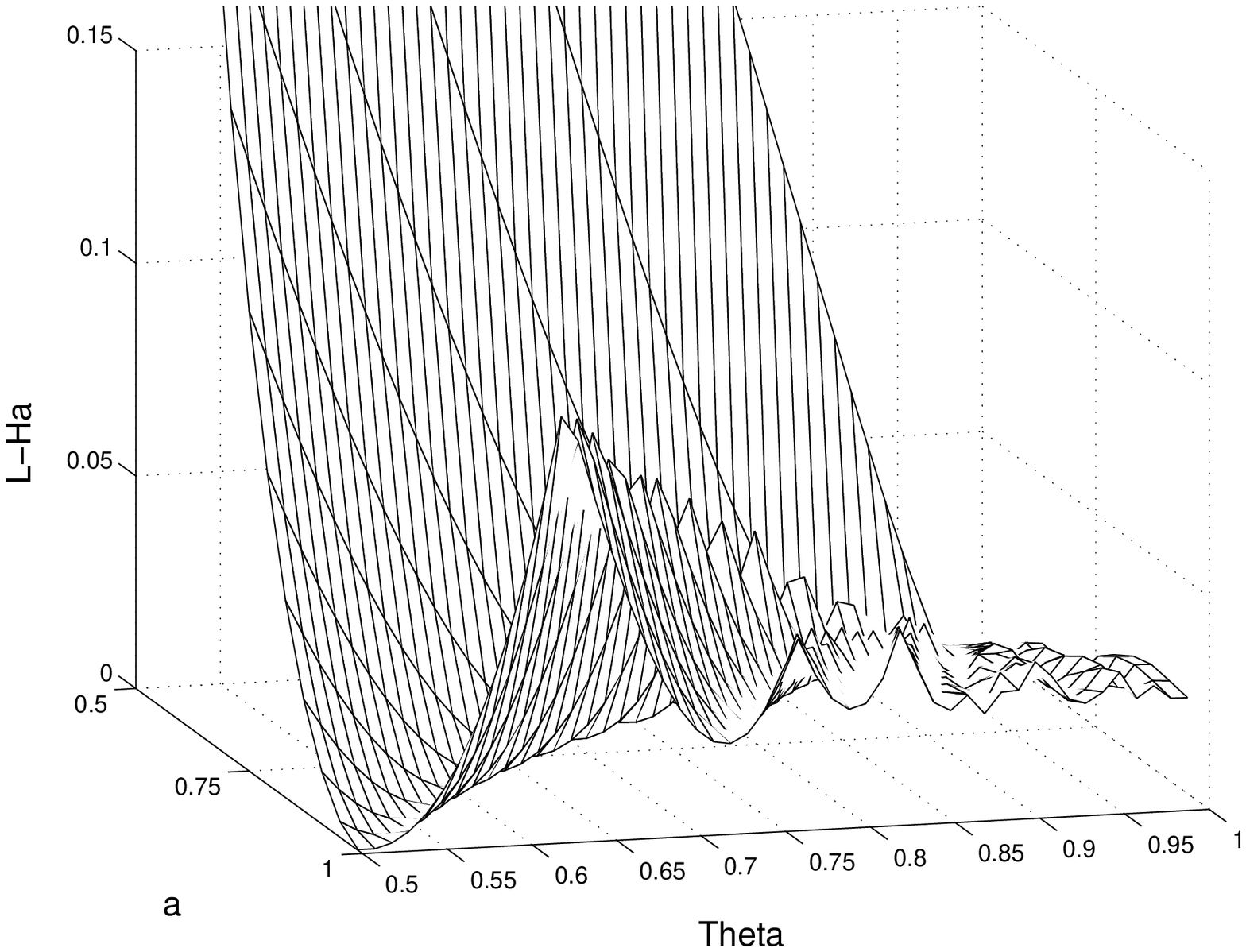} }
     \caption{Redundancy of the optimal code for the geometric
     distribution with the exponential penalty (parameter $a$).
     $\bar{R}_a(\biglen_{\theta,a}^*,\bigp_\theta) =
     \CampCost_a(\bigp_\theta,\biglen_{\theta,a}^*) - H_{\alpha(a)}(\bigp_\theta)$,
     where $\alpha(a) = (1+\lg a)^{-1}$, $\bigp_\theta$ is the
     geometric probability sequence implied by $\theta$, and
     $\biglen_{\theta,a}^*$ is the optimal length sequence for
     distribution $\bigp_\theta$ and parameter $a$.}
     \label{aall}
\end{figure*}

\begin{figure}[ht]
\psfrag{L-H}{\mbox{\huge $\bar{R}_1(\biglen_{\theta,1}^*,\bigp_\theta)$}}
\psfrag{THETA}{\mbox{\huge $\theta$}}
     \centering
     \resizebox{6.5cm}{!}{\includegraphics{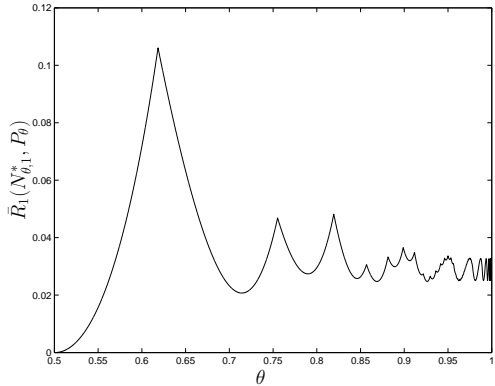}}
     \caption{Redundancy of the optimal code for the geometric
     distribution with the traditional linear penalty.}
     \label{shannon}
\end{figure}

In evaluating the effectiveness of the optimal code, one might use the
following definition of {\defn average pointwise redundancy} (or just
{\defn redundancy}): $$\bar{R}_a(\biglen, \bigp) \definedas
\CampCost_a (\bigp,\biglen) - H_{\alpha(a)}(\bigp) .$$ For
nondegenerate values, we can plot the $\bar{R}_a(\biglen_{\theta,a}^*,
\bigp_\theta)$ obtained from the minimization.  This is done for $a>1$
and $a<1$ in Fig.~\ref{aall}.  As $a \rightarrow 1$, the plot
approaches the redundancy plot for the linear case, e.g., \cite{GaVV},
reproduced as Fig.~\ref{shannon}.  In many potential applications of
nonlinear coding --- such as the aforementioned for
$a>1$\cite{Jeli,Humb2,BlMc} and $a<1$\cite{BaerI06,Baer07} --- $a$ is
very close to~$1$.  Since this analysis shows that the Golomb code
that is optimal for given $a$ and $\theta$ is optimal not only for
these particular values, but for a range of $a$ (fixing $\theta$) and
a range of $\theta$ (fixing $a$), the Golomb code is, in some sense,
much more robust and general than previously appreciated.

\section{Other Infinite Sources}
\label{other}

In this section we consider another type of probability distribution
for binary coding, a type with a light tail.  Humblet's
approach\cite{Humb1}, later extended in \cite{KHN}, uses the fact that
there is always an optimal code consisting of a finite number of
nonunary codewords for any probability distribution with a relatively
light tail, one for which there is an $r$ such that, for all $j>r$ and
$i<j$, $\boldp(i) \geq \boldp(j)$ and $\boldp(i) \geq
\sum_{k=j+1}^\infty \boldp(k)$.  Due to the additive nature of Huffman
coding, the unary part can be considered separately, and the remaining
codewords can be found via the Huffman algorithm.  Once again, this has to
be modified for the exponential case.

We wish to show that the optimal code can be obtained when there is a
nonnegative integer $r$ such that, for all $j>r$ and $i<j$, $$\boldp(i)
\geq \max\left(\boldp(j), \sum_{k=j+1}^\infty \boldp(k) a^{k-j}\right).$$

The optimal code is obtained by considering the reduced alphabet
consisting of symbols $0,1,\ldots,r+1$ with weights
\begin{equation}
\boldw(i) = \left\{
\begin{array}{ll}
\boldp(i),& i \leq r \\
\sum_{k=r+1}^\infty \boldp(k) a^{k-r},& i = r+1 . \\
\end{array}
\right.
\label{weights}
\end{equation}
Apply exponential Huffman coding to this reduced set of weights.  For
items $0$ through $r$, the Huffman codewords for the reduced and the
infinite alphabets are identical.  Each other item $i>r$ has a
codeword consisting of the reduced codeword for item $r+1$ (which,
without loss of generality, consists of all $1\s$) followed by the
unary code for $i-r-1$.  We call such codes {\defn unary-ended}.

\begin{theorem}
Let $\boldp(\cdot)$ be a probability measure on the set of nonnegative
integers, and let $a$ be the parameter of the penalty to be optimized.
If there is a nonnegative integer $r$
such that for all $j>r$ and $i<j$,
\begin{equation}
\boldp(i) \geq \max\left(\boldp(j),\sum_{k=j+1}^\infty \boldp(k)
a^{k-j}\right)
\label{cond}
\end{equation} then there exists a minimum-penalty binary prefix
code with every codeword~$j>r$ consisting of $j-x$ $1\s$ followed by
one $0$ for some fixed nonnegative integer~$x$.
\label{tailthm}
\end{theorem}

The proof of optimality (in full version \cite{Baer07}) is similar to
that for the geometric distribution.  In this case, for a given $m
\geq -1$, the corresponding codeword weights are
$$
\boldw_m(i) = \left\{
\begin{array}{ll}
\boldp(i),& i < i_{\max} \\
\sum_{k=i_{\max}}^\infty \boldp(k) a^{k-i_{\max}+1},& i = i_{\max} \\
\end{array}
\right.
$$ where $i_{\max} = r+m+2$.  For $a<1$, the proof is outlined
similarly to that for the geometric case.  For $a>1$, the key is to
note that the combined weight of a node in an optimal code is
upper-bounded by the weight of a node with the same children in a code
for which the node is the root of a unary subtree.  This allows an
inductive proof that the unary subtree --- and thus the proposed code
--- is optimal.

Consider the example of optimal codes for the Poisson distribution,
$$\boldp_\lambda(i)=\frac{\lambda^i e^{-\lambda}}{i!} . $$ How does
one find a suitable value for $r$ in such
a case?  It has been shown that $r \geq \lceil e \lambda \rceil - 1$
yields $\boldp(i) \geq \boldp(j)$ for all $j>r$ and $i<j$, satisfying
the first condition of Theorem~\ref{tailthm} \cite{Humb1}.  Moreover,
if, in addition, $j \geq \lceil 2 a \lambda \rceil - 1$ (and thus $j >
a \lambda - 1$), then
$$
\begin{array}{l}
\sum_{k=1}^\infty \boldp(j+k)a^k \\
\begin{array}{rcl}
&=& \frac{e^{-\lambda}\lambda^j}{j!}\left[
\frac{a \lambda}{j+1} + \frac{a^2 \lambda^2}{(j+1)(j+2)} + \cdots \right] \\
\qquad &<& \boldp(j) \left[\frac{a \lambda}{j+1} + \frac{a^2 \lambda^2}{(j+1)^2} + \cdots \right] \\
\qquad &=& \boldp(j) \frac{\frac{a \lambda}{j+1}}{1-\frac{a \lambda}{j+1}} \\
\qquad &\leq& \boldp(j) \\
\qquad &\leq& \boldp(i) .
\end{array}
\end{array}
$$
Thus, since we consider $j > r$, $r = \max(\lceil 2 a \lambda \rceil -
2, \lceil e \lambda \rceil - 1)$ is sufficient to establish an $r$
such that the above method yields the optimal infinite-alphabet code.

In order to find the optimal reduced code, use
\begin{eqnarray*}
\boldw_{-1}(r+1)&=&\sum_{k=r+1}^\infty \boldp(k) a^{k-r} \\
&=& a^{-r}e^{\lambda(a-1)} - \sum_{k=0}^r \boldp(k) a^{k-r} .
\end{eqnarray*}
For example, consider the Poisson distribution with $\lambda = 1$.  We
code this for both $a=1$ and $a=2$.  For both values, $r = 2$, so both
are easy to code.  For $a=1$, $\boldw_{-1}(3) = 1 - 2.5 e^{-1} \approx
0.0803 \ldots$, while, for $a=2$, $\boldw_{-1}(3) = 0.25 e - 1.25
e^{-1} \approx 0.2197 \ldots$.  After using the appropriate Huffman
procedure on each reduced source of $4$ weights, we find that the
optimal code for $a=1$ has lengths $\biglen = \{1, 2, 3, 4, 5, 6, \ldots\}$
--- those of the unary code --- while the optimal code for $a=2$ has
lengths $\biglen = \{2, 2, 2, 3, 4, 5, \ldots\}$.

\section{Redundancy penalties}
\label{nonexp}

It is natural to ask whether the above results can be extended to
other penalties.  One penalty discussed in the literature is that of
maximal pointwise redundancy\cite{DrSz}, in which one seeks to find
a code to minimize
$$R^*(\biglen,\bigp) \definedas \max_{i \in \X} [\len(i)+\lg
\boldp(i)].$$ 

This can be shown to be a limit of the exponential case, as in
\cite{Baer05}, allowing us to analyze it using the same techniques as
exponential Huffman coding.  This limit can be shown by defining
{\defn $d$th exponential redundancy} as follows:
$$
\begin{array}{rcll}
R_d(\biglen,\bigp) &\definedas&
\frac{1}{d} \lg \sum_{i \in \X} \boldp(i) 2^{d\left(\len(i)+\lg \boldp(i)\right)} \\
 &=& \frac{1}{d} \lg
\sum_{i \in \X} \boldp(i)^{1+d} 2^{d\len(i)}.
\end{array}
$$ Thus $R^*(\biglen,\bigp) = \lim_{d \rightarrow \infty}
R_d(\biglen,\bigp)$, and the above methods should apply in the limit.
In particular, the Golomb code G$\kval$ for $\kval = \lceil -1/\lg
\theta \rceil$ is optimal for minimizing maximum pointwise redundancy
for $\bigp_\theta$.  For light tails, a similar condition to
(\ref{cond}) holds; in this case, we find an $r$ such that,
$$\mbox{for all } i<r,~\boldp(i) \geq \boldp(r)$$
and 
$$\mbox{for all } j \geq r,~\boldp(j) \geq 2 \boldp(j+1).$$ 
Applications and proofs of these results are in the full version \cite{Baer07}.

\ifx \cyr \undefined \let \cyr = \relax \fi

\end{document}